\documentclass{aastex701}
\usepackage{amsmath,amssymb,bm}
\usepackage{graphicx}
\usepackage{xcolor}
\usepackage{enumitem}
\usepackage{hyperref}

\newcommand{\p}[1]{{\color{magenta}{#1}}}

\begin{document}


\shorttitle{Kinetic model of plasma atmospheres}
\shortauthors{Barbieri, D\'emoulin, \& Verscharen}

\title{Gravitational filtering, magnetic-field inhomogeneity and
Coulomb collisions in a kinetic model of a plasma atmosphere
with stochastic heating at its base}

\author{Luca Barbieri}
\affiliation{LIRA, Observatoire de Paris, Universit\'e PSL, CNRS, Sorbonne Universit\'e, Universit\'e Paris-Cit\'e, 5 place Jules Janssen, 92195 Meudon, France}
\email[show]{luca.barbieri@obspm.fr}
\correspondingauthor{Luca Barbieri}

\author{Pascal D\'emoulin}
\email{} 
\affiliation{LIRA, Observatoire de Paris, Universit\'e PSL, CNRS, Sorbonne Universit\'e, Universit\'e Paris-Cit\'e, 5 place Jules Janssen, 92195 Meudon, France}

\author{Daniel Verscharen}
\email{} 
\affiliation{Ruhr University Bochum, Faculty of Physics and Astronomy, Theoretical Physics I, Theoretical Plasma Physics, Bochum, Germany}
\affiliation{Mullard Space Science Laboratory, University College London, Holmbury St.~Mary, Dorking, United Kingdom}

\begin{abstract}
We investigate the combined effects of stochastic heating, magnetic inhomogeneity, and Coulomb collisions in coronal loops. Building on collisionless expanding-flux-tube models and constant-cross-section collisional models, we develop a kinetic model where Coulomb collisions both thermalize suprathermal populations from intermittent heating and populate velocity space disconnected from the lower boundary without collisions. We consider two heating-time-scale regimes. When heating is faster than the electron crossing time, the plasma is described by a multitemperature non-thermal distribution. When heating is slower than the proton crossing time, individual heated loops approach an approximately isothermal state. In the collisionless limit, magnetic-field inhomogeneity strongly modifies the accessible velocity space through conservation of the magnetic moment. This produces a pronounced temperature anisotropy, with parallel temperature larger than the perpendicular one, together with an additional depletion of the density. We find that Coulomb pitch-angle scattering efficiently suppresses these effects for typical coronal plasma parameters. In the short-time-scale regime, the density and temperature profiles become close to those obtained with a constant magnetic field, while gravitational and collisional filtering remain responsible for the temperature inversion. In the strongly collisional long-time-scale regime, the temperature anisotropy associated with magnetic-field inhomogeneity is almost completely suppressed, and individual heated loops remain approximately isotropic and isothermal. An increasing temperature profile emerges only after spatial averaging over an ensemble of unresolved flux tubes characterized by different temperatures. These results show that, under conditions representative of the transition region and the low corona, Coulomb collisions largely suppress the macroscopic kinetic signatures produced by magnetic field inhomogeneity.
\end{abstract}


\section{Introduction}
\label{sec:introduction}

\subsection{Coronal heating and kinetic transport}
\label{subsec:intro_kinetic}

The origin of the million-degree solar corona remains one of the major open questions in solar and plasma physics \citep{Aschwandensolarcorona,Klimchuk_2006,GolubPasachoff:book,Klimchuk2015}. Across the narrow transition region, the temperature of the solar atmosphere rises sharply from approximately $10^4$~K in the upper chromosphere to values of order $10^6$~K in the overlying corona,
while the density decreases by several orders of magnitude. Understanding the physical mechanisms responsible for supplying the energy required to maintain these extreme coronal temperatures remains the central challenge of the long-standing coronal heating problem.

A broad range of physical mechanisms has been proposed to account for the high temperature of the solar corona. One class of models invokes impulsive energy release associated with magnetic reconnection, including the nanoflare scenario originally proposed by \citet{Parker:1972wu}. In this picture, the continual stressing and reconfiguration of the coronal magnetic field lead to numerous small-scale energy-release events \citep{2005ApJ...618.1020G,Mason2023,Bose2024,Johnston_2025,
Mondal_2025}. Another class of coronal heating models invokes the dissipation of magnetohydrodynamic waves through mechanisms such as resonant absorption and phase mixing \citep{Ionson_1978,Heyvaerts_Priest_1983,VanDoorsselaere2020}. Other models emphasize the development of nonlinear interactions and turbulence, driven by waves or by the continuous stressing
of the magnetic field. These processes transfer energy toward smaller spatial scales, where dissipation can contribute to plasma heating
\citep{Dmitruk:1997uf,Chandran2005,Rappazzo:2008vl,
Dahlburg2012,2013ApJ...773L...2R}. The conversion of electromagnetic energy into particle energy is likely to involve processes operating at kinetic scales.
Wave--particle interactions and related kinetic mechanisms can transfer energy from electromagnetic fluctuations to particles, producing heating and generating non-thermal populations and
departures from Maxwellian velocity distribution functions
\citep{Vocks2001,Vocks_2002a,Vocks_2002b,Marsch2003,Chandran_2010,
Bian2010,Bian2011,Vocks2021,Ayaz_2024,Ayaz2025a,Ayaz2025b,
Ayaz2026a,Ayaz2026b}. Global magnetic models incorporating several of these physical processes have also been developed to investigate their combined role in the thermal structure of the solar atmosphere \citep{Chandran_2021,Yalim_2023,Baratashvili2024,Usmanov_2025,Baratashvili2026}.
 
A complementary question, central to the present work, is how the resulting particle populations subsequently redistribute this deposited energy and how their kinetic transport determines the macroscopic density and temperature profiles of the solar atmosphere. This question
is particularly relevant in the corona and transition region, where departures from local thermodynamic equilibrium and non-Maxwellian velocity distribution functions make a kinetic description necessary
\citep{Dudik:2017un,DelZanna2018,lazar2021kappa}. An important example of such a kinetic mechanism is velocity filtration or gravitational filtering. \citet{Scudder1992a,Scudder1992b} proposed that,
if the velocity distribution functions in the upper chromosphere possess suprathermal tails, particles with higher energies can reach greater heights in the atmosphere. The relative contribution of energetic particles therefore increases with
altitude, producing an increase in kinetic temperature accompanied by a decrease in density. This anti-correlation between the density and the temperature profiles is commonly referred to as temperature inversion. Velocity filtration does not provide an additional source of energy to the corona; rather, it describes how energy already deposited in the lower
atmosphere is redistributed through particle transport in a gravitationally stratified plasma. This mechanism has motivated subsequent kinetic descriptions of the solar atmosphere \citep{Meyer-Vernet_2007,Pierrard2014,Hau_2025,Banik2026}.

Similar kinetic approaches have also been widely employed to investigate the acceleration of the solar wind, the continuous outflow of charged particles from the solar atmosphere \citep{Parker1958}. These approaches, known as ``exospheric models'', describe the evolution of particle velocity distribution functions above a prescribed reference altitude by specifying their distributions at a lower boundary and following particle trajectories under the combined influence of gravitational, ambipolar electric field and magnetic-field inhomogeneity \citep{Jockers1970,Lemaire1971,Pierrard1996,Maksimovic1997,PIERRARD2001,Lamy2003,Zouganelis2004,Pierrard2008,PtersdeBonhome2025,Vinogradov2026}.

\subsection{Stochastic heating and velocity filtration}
\label{subsec:intro_stochastic_heating}

The present work shares this kinetic perspective and we consider the gravitationally stratified lower corona, where the plasma is structured by closed coronal magnetic
loops \citep[e.g.][]{Aschwandensolarcorona}. We therefore model a closed magnetic flux tube and investigate how particle transport along the loop determines its density and
temperature profiles. In a series of recent studies \citep{barbieri2024temperature, Barbieri2024b, barbieri2025temperaturedensityprofilescorona, Barbieri2025b, Barbieri2025c, barbieri2026, barbieri2026c, barbieri2026b}, we investigate this problem by connecting velocity filtration with intermittent energy deposition in the lower atmosphere. Observations reveal a large number of transient brightenings and energy-release events occurring over a broad range of spatial and temporal scales (see \cite{Young2018-sp,parenti2025,Harra2025} and references therein). A key difference with previous kinetic approaches is the origin of the particle distribution function at the lower boundary. While velocity-filtration models of the lower solar atmosphere prescribe the boundary distribution function \citep{Scudder1992a,Scudder1992b,Meyer-Vernet_2007,Pierrard2014,Hau_2025}, our approach derives it from the statistics of stochastic heating events. Heating events characterized by different intensities generate particle populations at different temperatures, whose superposition produces non-Maxwellian distributions with a suprathermal tails. The subsequent gravitational filtering of these populations then generates a transition from chromospheric to coronal temperatures.

Two different regimes arise depending on the characteristic time scales of the stochastic heating. When both the heating duration and the waiting time between successive heating events are shorter than the electron crossing time, populations generated at different temperatures coexist within the same flux tube. The resulting multi-temperature distribution remains out of
thermal equilibrium, and gravitational filtering produces temperature inversion. In contrast, when the heating time scale is comparable to or longer than the proton crossing time, each
heated flux tube has sufficient time to relax toward an approximately isothermal configuration. In this long-time-scale regime, the temperature inversion does not develop within an individual loop, but emerges after coarse graining of the surface over unresolved flux tubes
characterized by different temperatures \citep{Barbieri2025c}. The role of Coulomb collisions in these two regimes is investigated by \cite{barbieri2026}. In the short-time-scale regime, collisions progressively thermalize the suprathermal populations generated by the stochastic heating events. Dynamical friction and velocity-space diffusion reduce the fraction of energetic particles capable of reaching coronal heights, modifying the density profile while preserving
the qualitative temperature inversion. In the long-time-scale regime, by contrast, individual heated columns are already close to thermal equilibrium and are highly collisional. Coulomb collisions therefore produce only minor modifications of the approximately isothermal profiles of each heated flux tube. The temperature inversion obtained after surface coarse graining consequently remains unchanged.

\subsection{Magnetic field inhomogeneity and scope of the present work}
\label{subsec:intro_present_work}

These previous models, however, consider flux tubes with constant cross section. The magnetic field of the solar atmosphere is inherently inhomogeneous, and magnetic flux tubes generally expand with height as the magnetic-field strength decreases away from their photospheric footpoints \citep{Gabriel1976,Dowdy1986,Solanki_1991,Demoulin_1994,Mandrini_2000,Judge_2021}. In a magnetized plasma, this expansion introduces an additional kinetic effect through conservation of the magnetic moment. In our recent collisionless treatment of an expanding flux tube \citep{barbieri2026b}, we show that the conservation of particle energy and magnetic moment, together with the boundary connectivity condition, restricts the region of local velocity space populated by particles originating from the lower boundary. This pitch-angle-dependent restriction leads to a density depletion and a pronounced temperature anisotropy, with parallel and perpendicular temperatures shaping differently with height. These effects persist for both single-temperature and multi-temperature boundary distributions and can substantially modify the thermodynamic profiles predicted by gravitational filtering alone.

However, the collisionless approximation leaves open an important question. The same Coulomb collisions that thermalize suprathermal populations also scatter particles in pitch angle and therefore redistribute particles across the boundary separating the regions of velocity space that are connected and disconnected from the lower boundary in the collisionless mapping.  Magnetic field inhomogeneity and Coulomb
collisions thus act in competing directions: conservation of the magnetic moment tends to generate velocity-space anisotropy, whereas collisional scattering tends to isotropize the distribution and populate velocity-space regions that are inaccessible through collisionless transport alone. Whether the signatures of magnetic field inhomogeneity survive under the collisional conditions relevant to the transition region and low corona therefore cannot be established from either of our previous limiting models. The purpose of the present work is to address this problem by combining gravitational filtering, magnetic-field inhomogeneity and Coulomb collisions with stochastic heating at the base of the transition.

The paper is structured as follows. Section~\ref{sec:model_description} introduces the physical model and the basic equations. Section~\ref{sec:collisions_expanding_tube} develops the short-time-scale regime and investigates
the interplay between stochastic heating, magnetic field inhomogeneity, and Coulomb collisions. Section~\ref{sec:long_time_scale_regime} considers the long-time-scale regime. Finally, Section~\ref{sec:summary} summarizes the main results and outlines the limitations and
future developments of the model.

\section{Model overview}
\label{sec:model_description}

We consider a plasma atmosphere composed of electrons and protons that is globally at rest. The atmosphere is embedded in a closed, loop-like magnetic field. We further assume a symmetric magnetic loop, with equal plasma pressure at both footpoints, so that no net mass flow develops along the loop. The particle dynamics is described along the axis of a thin magnetic flux tube. We introduce the curvilinear coordinate $s$, measured along the magnetic-field line, and write the magnetic field as
\begin{equation}\label{eq:magneticfield}
    \mathbf{B}(s)
    =
    B(s)\,\mathbf{b}(s),
\end{equation}
where $\mathbf{b}(s)$ is the unit vector tangent to the magnetic-field line. The magnetic-field strength $B(s)$ is assumed to decrease monotonically from the loop footpoint along the portion of the flux tube considered here. Each particle is subjected to gravity,
\begin{equation}
    \mathbf{F}_{g,\alpha}
    =
    m_{\alpha}\, \mathbf{g},
    \qquad
    \alpha\in\{e,p\},
\end{equation}
where $g$  is the gravitational acceleration at the solar surface, while $m_e$ and $m_p$ denote the electron and proton masses. The plasma is also subjected to the Pannekoek--Rosseland electric field \citep{Pannekoek_1922,Rosseland_1924,Neslusan2001-rp,Barbieri_2025c},
\begin{equation}
    \label{Pannekoekfield}
    \mathbf{E}_{\mathrm{PR}}
    =
    -\frac{m_p-m_e}{2\,e}\,\mathbf{g}.
\end{equation}
Under these assumptions the total force acting on a particle of species $\alpha$ is therefore
\begin{equation}
    \label{totalforce}
    \mathbf{F}_{\alpha}
    =
    \overline{m}\, \mathbf{g}
    +
    \frac{e_{\alpha}}{c}
    \mathbf{v}\times\mathbf{B}(s)
    \qquad \mathrm{with} \;
    \qquad\overline{m}
    =
    \frac{m_e+m_p}{2},
\end{equation}
where $e_{\alpha}=e$ for protons, $e_{\alpha}=-e$ for electrons and $c$ is the speed of light. Since the dynamics is described along a curved magnetic-field line, the component of the total force along the flux tube is described by the corresponding potential energy $\phi(s)$ through
\begin{equation}
    -\frac{{\rm d}\phi}{{\rm d}s} = F_{\alpha,\parallel}(s)
    =
    \overline{m} \,\mathbf{g}\cdot\mathbf{b}(s).
    \label{eq:effective_potential_curvilinear}
\end{equation}
The corresponding single-particle Hamiltonian is
\begin{equation}
    H_\alpha(s,\mathbf{v})
    =
    \frac{1}{2}m_\alpha v^2
    +
    \phi(s).
    \label{eq:single_particle_hamiltonian}
\end{equation}

In the following, the equations apply to both protons and electrons upon specifying the corresponding particle properties. For conciseness, we omit the species index $\alpha$.

\section{Fast time-scale regime}
\label{sec:collisions_expanding_tube}

We now extend the collisional model introduced by \cite{barbieri2026} for a constant-cross-section magnetic flux tube to the expanding geometry considered here. As in the collisionless model \citep{barbieri2026b}, conservation of particle energy and magnetic moment divides the local velocity space into regions that are connected and disconnected from the lower boundary through collisionless Liouville mapping. Coulomb collisions are incorporated through a reduced closure accounting for two effects: the progressive attenuation of the suprathermal populations and the collisional filling of the disconnected velocity-space region through pitch-angle scattering. These processes are described by the functions $\omega(T,s)$ and $R(T,s)$ as detailed below.

\subsection{Boundary conditions}
\label{subsec:boundary_condition}

At the lower boundary of the magnetic flux tube, $s=0$, the plasma is assumed to be in thermal contact with the upper chromosphere, which is modelled as a thermal boundary. In the absence of heating, particles are injected with a Maxwellian distribution at the background temperature $T_0=10^4 \mathrm{K}$. Following \citet{barbieri2026}, we assume that the temperature of the boundary undergoes stochastic fluctuations associated with intermittent heating events. During a heating event of characteristic duration $\tau$, the boundary temperature increases from $T_0$ to a temperature $T>T_0$, drawn from a probability distribution $\gamma(T)$. After the heating phase, the boundary returns to the background temperature $T_0$ for a characteristic waiting time $t_w$, after which a new heating event occurs. Presently, we restrict the analysis to the short-time-scale regime introduced in \citet{barbieri2024temperature} as
\begin{equation}
    \tau,t_w < t_{c,e},
    \label{eq:short_timescale_condition}
\end{equation}
where $t_{c,e}$ is the characteristic electron crossing time along the magnetic loop. In this regime, the boundary temperature fluctuates on time scales shorter than those required for particles to cross the entire system. Particle populations generated during successive heating and cooling phases therefore coexist and mix and the system reaches a non-thermal stationary state. The resulting stationary distribution function at the lower boundary can be written as
\begin{equation}
f(0,\mathbf{v}) =
N \bigg[
A_t \int_{T_0}^{+\infty} {\rm d}T\,
\frac{\gamma(T)}{T^{3/2}}
\exp\left(-\frac{m v^2}{2 k_{\rm B} T}\right)
+
\frac{1-A_t}{T_0^{3/2}}
\exp\left(-\frac{m v^2}{2 k_{\rm B} T_0}\right)
\bigg].
\label{eq:boundary_distribution}
\end{equation}
where
\begin{equation}
    A_t
    =
    \frac{\tau}{\tau+t_w},
    \label{eq:temporal_filling_factor}
\end{equation}
is the fraction of time during which heating is active. The function $\gamma(T)$ is the probability density of the temperatures reached during the heating phases, normalized such that
\begin{equation}
    \int_{T_0}^{+\infty}
    \gamma(T)\,{\rm d}T
    =
    1.
\end{equation}

The normalization constant $N$ is determined by requiring the density of each species at the lower boundary to be equal to $n_0$. Therefore,
\begin{equation}
    N
    =
    n_0
    \left(
    \frac{m}
    {2\pi k_{\rm B}}
    \right)^{3/2}.
    \label{eq:N_alpha_normalization}
\end{equation}

Equation~\eqref{eq:boundary_distribution} consists of a cold Maxwellian population at the chromospheric temperature $T_0$, with relative weight $1-A_t$, and a suprathermal component produced by stochastic heating, represented by a continuous superposition of Maxwellian populations at $T>T_0$ with total relative weight $A_t$. The latter provides the suprathermal population that is subsequently modified by gravitational
filtering, magnetic field inhomogeneity, and Coulomb collisions as particles propagate along the flux tube. As discussed by \citet{barbieri2026}, describing each individual heating event by a Maxwellian distribution is a conservative approximation. In the short-time-scale regime, the duration of the heating events may be comparable to, or shorter than, the local collisional relaxation time at the base of the transition region, so that complete thermalization at the boundary is not necessarily achieved. More realistic heating mechanisms could generate intrinsically non-Maxwellian distributions with enhanced suprathermal tails. The Maxwellian assumption adopted here minimizes this non-thermal contribution.

\subsection{Passing and trapped velocity-space domain}
\label{subsec:passing_domains}

As discussed in the collisionless model by \citet{barbieri2026b}, as well as in exospheric models of the solar wind \citep{Jockers1970,Lemaire1971,Pierrard1996,Maksimovic1997,PIERRARD2001,Lamy2003,Zouganelis2004,Pierrard2008,PtersdeBonhome2025,Vinogradov2026}, conservation of the single-particle energy and magnetic moment divides the local velocity space into a domain directly connected to the lower boundary and a complementary trapped region. We define the characteristic function  $\chi(s,\mathbf{v})$ with $\chi=1$ for trajectories that can be mapped collisionlessly to the lower boundary and $\chi=0$ otherwise. We recall the quantities defining the boundary between the two velocity-space domains. We introduce
\begin{equation}
    c(s)
    =
    \sqrt{
    \frac{B(0)-B(s)}
    {B(0)}
    },
    \label{eq:c_definition}
\end{equation}
which measures the magnetic field inhomogeneity between the lower boundary and the position $s$. The critical velocity separating the two angular regimes is
\begin{equation}
    v_{B}^2(s)
    =
    \frac{B(s)}
    {B(0)-B(s)}
    \frac{2\,\phi(s)}{m}.
    \label{eq:vB_definition}
\end{equation}
with $\phi(s)$ computed by integration of Equation \eqref{eq:effective_potential_curvilinear}. It is also convenient to introduce the dimensionless quantity
\begin{equation}
    a^2(T,s)
    =
    \frac{m\, v_{B}^2(s)}
    {2\,k_{\rm B}\, T}.
    \label{eq:a_definition}
\end{equation}
The parameter $a$ expresses the critical velocity $v_{B}$ in units of the thermal velocity of the Maxwellian population generated at temperature $T$. For $v\leq v_{B}(s)$, all pitch angles belong to the velocity-space domain connected to the lower boundary. For $v>v_{B}(s)$, instead, only particles satisfying
\begin{equation}
    |\theta|\leq\theta_m(s,v)
\end{equation}
remain collisionlessly connected to the lower boundary, where the limiting pitch angle is defined by
\begin{equation}
    \cos\theta_{m}(s,v)
    =
    c(s)
    \frac{
    \sqrt{v^2-v_{B}^2(s)}
    }{v}.
    \label{eq:theta_m_definition}
\end{equation}

Therefore the characteristic function $\chi$ is given
\begin{equation}
\label{velocityfilter}
\chi(z,\mathbf v)=
\begin{cases}
1, & v \le v_{B}(z), \\[3pt]
1, & v>v_{B}(z), \,|\theta|\le\theta_m(z,v), \\[3pt]
0, & v>v_{B}(z), \,|\theta|>\theta_m(z,v).
\end{cases}
\end{equation}

\subsection{Collisional distribution function}
\label{subsec:collisional_distribution}

We decompose the collisional distribution function $f$ into a multitemperature suprathermal component $f_H$ and a Maxwellian core $f_M$,
\begin{equation}
    f = f_H +f_M\,.
    \label{eq:collisional_decomposition}
\end{equation}

\subsubsection{Suprathermal component and collisional filtering}
\label{subsubsec:suprathermal_filtering}

With the boundary conditions set by Equation~\eqref{eq:boundary_distribution}, the suprathermal contribution is written as
\begin{equation}
f_H(s,\mathbf{v}) =
N A_t \int_{T_0}^{+\infty} {\rm d}T\,
\frac{\gamma(T)\,\omega(T,s)}{T^{3/2}}
\exp\left[-\frac{H(s,\mathbf{v})}{k_{\rm B}T}\right]
\left[\chi(s,\mathbf{v})+
(1-R(T,s))(1-\chi(s,\mathbf{v}))\right].
\label{eq:halo_distribution_collisional}
\end{equation}
Here, $\omega(T,s)$ and $R(T,s)$ are two collisional factors. We describe $\omega(T,s)$ in this subsection and then $R(T,s)$ in subsection \ref{subsubsec:trapped_filling}. The factor $\omega(T,s)$ describes the progressive attenuation, due to Coulomb collisions, of the contribution associated with the Maxwellian population initially injected at temperature $T$ at the lower boundary. Its evolution along the field line is modelled by
\begin{equation}
    \frac{\partial \omega(T,s)}{\partial s}
    =
    -\frac{\omega(T,s)}{\lambda(T,s)},
    \label{eq:omega_evolution}
\end{equation}
where $\lambda(T,s)$ is the collisional mean free path of the population at temperature $T$. $\omega(T,0)=1$ is the boundary condition at $s=0$. Equation~\eqref{eq:omega_evolution} was derived from the kinetic transport arguments presented in detail in \citet{barbieri2026}. Although that study considered a flux tube with constant cross section, the derivation of this evolution equation does not rely on that geometrical assumption. We therefore retain the same evolution equation in the present model, while evaluating the mean free path $\lambda(T,s)$ self-consistently from the moments of the distribution function, taking into account the local plasma conditions and the expanding
magnetic-field geometry (see Section~\ref{subsec:collisional_mean_free_path}below).

Thus, $\omega(T,s)$ represents the surviving fraction of the original population associated with temperature $T$ after collisional processing. Populations characterized by a short mean free path are rapidly attenuated by Coulomb collisions, whereas those with a sufficiently long mean free path are less affected by collisional relaxation and remain approximately described by the collisionless mapping of the distribution prescribed at the lower boundary. This attenuation does not represent a loss of particles from the system. Rather, particles removed from the
suprathermal population by collisions are incorporated into the Maxwellian core (see Section~\ref{subsec:collisional_mean_free_path}below).

\subsubsection{Collisional filling of the trapped domain}
\label{subsubsec:trapped_filling}

The second collisional factor appearing in
Equation~\eqref{eq:halo_distribution_collisional} is $R(T,s)$, which describes the progressive filling, as a function of $s$, of the velocity-space domain that is not collisionlessly connected to the lower boundary. Coulomb collisions, however, redistribute particles in velocity space and transfer particles from the connected domain into the otherwise inaccessible domain. We therefore introduce the factor $1-R(T,s)$ as a measure of its degree of filling: $R=1$ corresponds to the collisionless limit, in which this domain is empty, whereas $R=0$ corresponds to complete filling with the Maxwellian population at temperature $T$. In the present model, we assume that the collisional filling of the disconnected velocity-space domain occurs over the same collisional length $\lambda(T,s)$ introduced above for the survival factor $\omega(T,s)$. Accordingly, $R(T,s)$ obeys
\begin{equation}
\frac{\partial R(T,s)}{\partial s}
=
-\frac{R(T,s)}{\lambda(T,s)},
\qquad
R(T,0)=1.
\label{eq:R_evolution}
\end{equation}

The boundary condition $R(T,0)=1$ follows from the injection condition at the lower boundary. At $s=0$, $c(0)=0$, so that the disconnected velocity-space domain has zero volume and $\chi=1$  everywhere. Thus, $R(T,0)=1$ specifies that no collisional filling of a disconnected domain has yet accumulated for the population at the point of injection. Although Equation~\eqref{eq:R_evolution} has the same mathematical form as Equation~\eqref{eq:omega_evolution}, the two factors have distinct physical meanings, as discussed above. Unlike the evolution equation for $\omega(T,s)$, which was derived in \citet{barbieri2026}, adopting the same equation for $R(T,s)$ is a phenomenological assumption introduced to describe the collisional filling of the trapped velocity-space domain by particles associated with different temperature components.
Within this prescription, hotter components, characterized by longer mean free paths, tend to retain the collisionless boundary-connected distribution, whereas colder components undergo stronger collisional redistribution, leading to fill more the trapped domain. The limitations of this assumption are discussed in Section~\ref{sec:summary}.

\subsubsection{Maxwellian core}
\label{subsubsec:maxwellian_core}

The particles removed from the suprathermal populations through collisional thermalization are transferred to the Maxwellian core, whose distribution function is
\begin{equation}
f_M(s,\mathbf{v}) =
N \frac{C(s)}{T_M^{3/2}(s)}
\exp\left[-\frac{H(s,\mathbf{v})}{k_{\rm B}T_M(s)}\right]
\left[\chi(s,\mathbf{v})+
(1-R_M(s))(1-\chi(s,\mathbf{v}))\right].
\label{eq:core_distribution_collisional}
\end{equation}
Here, $C(s)$ determines the amplitude of the Maxwellian core, $T_M(s)$ its temperature, and $R_M(s)$ describes the collisional filling of the velocity-space domain that is disconnected from the lower boundary through collisionless trajectories. At the lower boundary, the core must recover the Maxwellian component of the prescribed distribution in Equation~\eqref{eq:boundary_distribution}. We therefore impose
\begin{equation}
    T_M(0)=T_0,
    \qquad
    C(0)=1-A_t.
    \label{eq:core_boundary_conditions}
\end{equation}
At $s=0$, the collisionless disconnected velocity-space domain has zero volume and $\chi=1$. The core distribution therefore reduces to
\begin{equation}
    f_M(0,\mathbf{v})
    =
    N\,
    \frac{1-A_t}{T_0^{3/2}}
    \exp\left[
        -\frac{m v^2}{2k_{\rm B}T_0}
    \right],
    \label{eq:core_boundary_distribution}
\end{equation}
which is the Maxwellian contribution prescribed in
Equation~\eqref{eq:boundary_distribution}. As height increases, collisions between core and suprathermal particles result in the exchange of particles and energy between the two populations, thereby modifying the core amplitude $C(s)$ and its characteristic temperature $T_M(s)$ from their respective values at the base, $1-A_t$ and $T_0$. The collisional filling of the disconnected velocity-space domain is described for the Maxwellian core using the same reduced prescription adopted for the each temperature contribution in the suprathermal population. Therefore, the corresponding $R_M(s)$ obeys
\begin{equation}
    \frac{\partial R_M(s)}{\partial s}
    =
    -\frac{R_M(s)}
           {\lambda\!\left(T_M(s),s\right)},
    \qquad
    R_M(0)=1,
    \label{eq:RM_evolution}
\end{equation}
and the same physical considerations discussed for $R(T,s)$ in Section~\ref{subsubsec:trapped_filling} also apply to $R_M(s)$.

\subsection{Velocity moments}
\label{subsec:collisional_velocity_moments}

We obtain the kinetic moments of the collisional distribution function
defined in Equation~\eqref{eq:collisional_decomposition} by integrating separately the suprathermal and Maxwellian-core components over velocity
space. The velocity integrals involve the collisionlessly connected and trapped domains introduced in Section~\ref{subsec:passing_domains}. The detailed procedure for calculating the contribution from the collisionlessly connected domain is analogous to that presented by \citet{barbieri2026b}. The same procedure can be applied to the trapped domain by performing the integration over the complementary region of phase space. The number density of the complete distribution is
\begin{equation}
n(s) = n_0 \bigg[
A_t \int_{T_0}^{+\infty} {\rm d}T\,
\gamma(T)\,\omega(T,s)\,
e^{-\phi(s)/(k_{\rm B}T)}\,\mathcal{G}_*(T,s)
+ C(s)\,e^{-\phi(s)/[k_{\rm B}T_M(s)]}\,
\mathcal{G}_{*,M}(s)
\bigg].
\label{eq:collisional_density}
\end{equation}
The parallel and perpendicular pressures are
\begin{equation}
p_i(s) = n_0\,k_{\rm B} \bigg[
A_t \int_{T_0}^{+\infty} {\rm d}T\,
T\,\gamma(T)\,\omega(T,s)\,
e^{-\phi(s)/(k_{\rm B}T)}\,\mathcal{G}_i(T,s)
+ C(s)\,T_M(s)\,
e^{-\phi(s)/[k_{\rm B}T_M(s)]}\,
\mathcal{G}_{i,M}(s)
\bigg],
\qquad i\in\{\parallel,\perp\}.
\label{eq:collisional_parallel_pressure}
\end{equation}
The corresponding parallel and perpendicular temperatures are
\begin{equation}
T_i(s)
=
\frac{p_i(s)}
{k_{\rm B}\,n(s)},
\qquad
i\in\{\parallel,\perp\},
\label{eq:collisional_parallel_temperature}
\end{equation}
and the total temperature is
\begin{equation}
T(s)
=
\frac{
T_\parallel(s)+2T_\perp(s)
}{3}.
\label{eq:collisional_scalar_temperature}
\end{equation}
The dimensionless factors $\mathcal{G}_*$,
$\mathcal{G}_\parallel$, and $\mathcal{G}_\perp$ appearing in the density and pressure expressions account for the contributions of both the collisionlessly connected and trapped populations to the corresponding velocity moments. Since the trapped population is filled according to the factor $1-R(T,s)$, these complete geometrical factors are given by
\begin{equation}
\mathcal{G}_i(T,s) =
G_i(T,s) + [1-R(T,s)][1-G_i(T,s)]
= 1-R(T,s)[1-G_i(T,s)],
\qquad i\in\{*,\parallel,\perp\}.
\label{eq:G_completed_general}
\end{equation}
Here, $G_*$, $G_\parallel$, and $G_\perp$ denote, respectively, the normalized density, parallel-pressure, and perpendicular-pressure contributions of the collisionlessly connected population \citep{barbieri2026b}. Their explicit expressions are
\begin{equation}
G_*(T,s)
=
1-c(s)\,e^{-a^2},
\label{eq:G0}
\end{equation}
\begin{equation}
G_\parallel(T,s)
=
1-c^3(s)\,e^{-a^2},
\label{eq:Gparallel}
\end{equation}
\begin{equation}
G_\perp(T,s)
=
1-c(s)
\left[
\frac{3}{2}+a^2
\right]
e^{-a^2}
+
\frac{c^3(s)}{2}
e^{-a^2},
\label{eq:Gperp}
\end{equation}
where $c(s)$ and $a^2=a^2(T,s)$ are defined in
Equations~\eqref{eq:c_definition} and
\eqref{eq:a_definition}, respectively. For the Maxwellian core, the corresponding geometrical factors are obtained by evaluating the same expressions at the local core temperature $T_M(s)$, with the trapped
population filled according to the factor $1-R_M(s)$. The resulting complete core factors are denoted by $\mathcal{G}_{*,M}$, $\mathcal{G}_{\parallel,M}$, and $\mathcal{G}_{\perp,M}$. The limiting behaviour of these factors provides a direct connection between the collisionless and fully populated
velocity-space configurations. In the collisionless limit,
$R\rightarrow1$, $\mathcal{G}_i\rightarrow G_i$, and only the population collisionlessly connected to the lower boundary contributes to the moments. Conversely, in the limit of complete filling of the trapped domain,
$R\rightarrow0$, $\mathcal{G}_i\rightarrow1$,
and the full local Maxwellian moments are recovered.

\subsection{Collisional mean free path}
\label{subsec:collisional_mean_free_path}

The evolution of the collisional factors, $\mathcal{G}_*$, $\mathcal{G}_\parallel$, and $\mathcal{G}_\perp$, introduced above is controlled by the mean free path $\lambda(T,s)$. Since the collision rate depends on both the characteristic particle velocity and the local plasma state, $\lambda(T,s)$ generally depends on the distribution function itself. Following the prescription adopted in our constant-cross-section collisional model \citep{barbieri2026}, the mean free path associated with the Maxwellian population generated at temperature $T$ is 
\begin{equation}
\lambda(T,s)
=
\frac{
m^2
\left[
\left\langle v_{\parallel}^2\right\rangle_T
+
2\left\langle v_{\perp}^2\right\rangle_T
\right]^2
}{
4\pi\,e^4\,n(s)\,\ln\Lambda(T,s)
}.
\label{eq:halo_mean_free_path}
\end{equation}
Here, $v_\perp$ denotes one of the two perpendicular velocity components. The factor of $2$ in Equation~\eqref{eq:halo_mean_free_path} accounts for the two perpendicular degrees of freedom. The velocity moments are evaluated over the actual velocity-space population associated with temperature $T$. Using the geometric factors introduced above, we write
\begin{equation}
\left\langle v_{i}^2\right\rangle_T
=
\frac{k_{\rm B}\, T}{m}
\frac{\mathcal{G}_i(T,s)}
{\mathcal{G}_*(T,s)}, \quad
i\in\{\parallel,\perp\}.
\label{eq:beam_parallel_second_moment}
\end{equation}
The Coulomb logarithm is evaluated using the local density and temperature of the complete plasma distribution
\begin{equation}
\ln\Lambda(T,s)
=
\ln\left[
\frac{3\,k_{\rm B}\, T}{2\,e^2}
\sqrt{
\frac{k_{\rm B}\, T(s)}
{4\pi\,e^2\,n(s)}
}
\right].
\label{eq:coulomb_logarithm}
\end{equation}
The dependence of the mean-free-path on the geometric factors $\mathcal{G}_*$, $\mathcal{G}_\parallel$, and $\mathcal{G}_\perp$ incorporates the modification of the velocity moments produced by the inhomogeneous magnetic geometry and by the collisional filling of the trapped velocity-space domain.

\subsection{Evolution of the Maxwellian-core amplitude}
\label{subsec:core_amplitude}

As in the constant-cross-section model of \cite{barbieri2026}, the amplitude $C(s)$ of the Maxwellian core is determined by imposing local particle conservation during the collisional redistribution over a small distance $\delta s$. We therefore impose
\begin{equation}
    \left.
    \frac{\delta n(s)}{\delta s}
    \right|_{\rm coll}
    \approx
    \left.
    \frac{d n(s)}{d s}
    \right|_{\rm coll}
    =0,
    \label{eq:variationdensity}
\end{equation}
where the subscript ``coll'' indicates that only the variation induced by collisional redistribution is considered. The local macroscopic plasma quantities and magnetic geometry are assumed to remain constant. In particular, exponential factors associated with
gravitational stratification are treated as constant, since the characteristic length scale considered here is the Debye length, which is much smaller than the gravitational scale height. The collisionless geometrical factors $G_i(T,s)$ are also held fixed, whereas the functions describing collisional effects, namely $\omega(T,s)$, $R(T,s)$,
$R_M(s)$, and $C(s)$, are allowed to vary. This local approximation relies on the separation between the microscopic length scale considered here and the macroscopic scales over which the plasma properties and
magnetic geometry evolve. Using Equation~\eqref{eq:collisional_density}, with the definition of $\mathcal{G}_*(T,s)$ given by Equation~\eqref{eq:G_completed_general}, the corresponding density variation is
\begin{align}
\left.\frac{dn(s)}{ds}\right|_{\rm coll}
={}& A_t \int_{T_0}^{+\infty} {\rm d}T\,\gamma(T)\,
e^{-\phi/(k_{\rm B}T)}
\left[\mathcal{G}_*(T,s)\frac{\partial\omega(T,s)}{\partial s}
-\omega(T,s)[1-G_*(T,s)]\frac{\partial R(T,s)}{\partial s}\right]
\nonumber\\
&+e^{-\phi/[k_{\rm B}T_M(s)]}
\left[\mathcal{G}_{*,M}(s)\frac{dC(s)}{ds}
-C(s)[1-G_{*,M}(s)]\frac{dR_M(s)}{ds}\right].
\label{eq:density_collisional_variation}
\end{align}

Imposing Equation~\eqref{eq:variationdensity} together with Eqs. \eqref{eq:omega_evolution},\eqref{eq:R_evolution},\eqref{eq:RM_evolution} we get
\begin{align}
\frac{dC}{ds}
={}&
\frac{A_t}{e^{-\phi(s)/[k_{\rm B}T_M(s)]}\,\mathcal{G}_{*,M}(s)}
\int_{T_0}^{+\infty} {\rm d}T\,\gamma(T)\,
e^{-\phi(s)/(k_{\rm B}T)}\,\omega(T,s)
\left[
\frac{\mathcal{G}_*(T,s)}{\lambda(T,s)}
-\frac{R(T,s)[1-G_*(T,s)]}{\lambda(T,s)}
\right]
\nonumber\\
&-
\frac{C(s)R_M(s)[1-G_{*,M}(s)]}
{\mathcal{G}_{*,M}(s)\lambda_M(s)}.
\label{eq:C_evolution}
\end{align}

Equation~\eqref{eq:C_evolution} expresses local particle conservation during the collisional redistribution and makes explicit the coupling between collisional filtering, trapped-domain filling, and the evolution of the Maxwellian-core amplitude $C$. A detailed description of each term in Eq.~\eqref{eq:C_evolution} is provided in Appendix~\ref{app:C_terms}.

\subsection{Evolution of the Maxwellian-core temperature}
\label{subsec:core_temperature}

The Maxwellian-core temperature $T_M(s)$ is determined by requiring the complete particle distribution to satisfy stationary momentum balance along the magnetic flux tube. Since the velocity distribution is anisotropic but gyrotropic, its pressure tensor is
\begin{equation}
\mathbf{P}
=
p_{\perp}\mathbf{I}
+
\left(
p_{\parallel}-p_{\perp}
\right)
\mathbf{b}\otimes\mathbf{b}.
\label{eq:pressure_tensor_anysotropy}
\end{equation}

In the absence of a bulk flow, the stationary momentum equation is
\begin{equation}
\boldsymbol{\nabla}\cdot\mathbf{P}
+
n\,\boldsymbol{\nabla}\phi
=
0.
\end{equation}
Projecting this equation along the magnetic-field direction and using Maxwell's equation $\boldsymbol{\nabla}\cdot\mathbf{B}=0$,
we obtain
\begin{equation}
\frac{{\rm d}p_{\parallel}}{{\rm d}s}
-
\left(
p_{\parallel}-p_{\perp}
\right)
\frac{{\rm d}\ln B}{{\rm d}s}
=
-n(s)\frac{{\rm d}\phi}{{\rm d}s}.
\label{eq:anisotropic_pressure_balance}
\end{equation}

The first term in Equation~\eqref{eq:anisotropic_pressure_balance}
describes the variation of the parallel pressure along the flux tube, whereas the right-hand side represents the force density associated with the combined gravitational and Pannekoek potential.
The additional term, proportional to $p_{\parallel}-p_{\perp}$, arises from the coupling between pressure anisotropy and magnetic-field inhomogeneity. It vanishes when the pressure is
isotropic ($p_{\parallel}=p_{\perp}$) or when the magnetic-field strength is constant along the flux tube. Substituting Equations~\eqref{eq:collisional_density}--\eqref{eq:collisional_parallel_pressure} into Equation~\eqref{eq:anisotropic_pressure_balance}, differentiating the parallel pressure, and isolating ${\rm d}T_M/{\rm d}s$, we obtain
\begin{align}
\frac{dT_M}{ds}
={}&
\bigg\{
-A_t\frac{d\mathcal{H}_\parallel}{ds}
-\frac{1}{k_{\rm B}}\frac{d\phi}{ds}
\left(A_t\mathcal{H}_*+C E_M\mathcal{G}_{*,M}\right)
+\Delta\mathcal{P}\frac{d\ln B}{ds}
-\frac{dC}{ds}T_M E_M\mathcal{G}_{\parallel,M}
\nonumber\\
&\quad
-C T_M E_M\left[
\left.\frac{d\mathcal{G}_{\parallel,M}}{ds}\right|_{T_M}
-\frac{\mathcal{G}_{\parallel,M}}{k_{\rm B}T_M}\frac{d\phi}{ds}
\right]
\bigg\}
\bigg\{
C E_M\left[
\mathcal{G}_{\parallel,M}
\left(1+\frac{\phi}{k_{\rm B}T_M}\right)
+T_M\frac{d\mathcal{G}_{\parallel,M}}{dT_M}
\right]
\bigg\}^{-1}.
\label{eq:TM_evolution_compact}
\end{align}
The terms involved in Equation~\eqref{eq:TM_evolution_compact} are analysed in Appendix~\ref{app:TM_terms}. 
In the isotropic limit, $\Delta\mathcal{P}\rightarrow0$,
the mirror-force contribution disappears. In the constant-cross-section limit, ${\rm d}B / {\rm d}s \rightarrow 0$ the geometrical terms also vanishes and the equation reduces to the constant-cross-section collisional formulation of \cite{barbieri2026}. Equations \eqref{eq:omega_evolution}, \eqref{eq:C_evolution}, and \eqref{eq:TM_evolution_compact} form a closed nonlinear integro-differential system. The numerical integration follows the strategy introduced by \cite{barbieri2026}. We integrate the system along the curvilinear coordinate $s$ using an adaptive implicit algorithm, while all local plasma quantities and collisional mean free paths are recomputed self-consistently at each integration step. For further details on the numerical scheme, see Appendix~\ref{app:numerical_scheme}.

\subsection{Density and temperature profiles}
\label{subsec:single_profiles}

\subsubsection{Magnetic-field geometry and stochastic-heating distribution}
\label{subsubsec:vertical_geometry}

We now illustrate the density and temperature profiles for a representative inhomogeneous magnetic field. While the framework developed above is formulated in terms of the general curvilinear coordinate $s$ along the magnetic field, for the numerical examples presented in this subsection, we simplify our configuration to a vertical magnetic field and identify $s=z$. This simplification follows the approach adopted in our previous collisionless treatment \citep{barbieri2026b}. It is appropriate for the nearly vertical legs of sufficiently large magnetic loops and avoids introducing additional geometrical effects associated with the curvature of the field line. The purpose of the present calculation is therefore not to model the complete geometry of a closed magnetic loop, but to isolate the effects of magnetic-field inhomogeneity and Coulomb collisions on the vertical density and temperature profiles. Within this geometry, Equation~\eqref{eq:effective_potential_curvilinear} reduces to
\begin{equation}
    \frac{d\phi}{dz}=\overline{m}\,g,
\end{equation}
where $g$ is taken to be constant over the spatial domain considered. Choosing $\phi(0)=0$, the potential therefore takes the standard linear form
\begin{equation}
    \phi(z)=\overline{m}\,g\,z.
    \label{eq:effective_potential_cartesian}
\end{equation}

For consistency with the collisionless analysis of \citet{barbieri2026b}, we adopt the same monotonically decreasing magnetic-field profile,
\begin{equation}
B(z)
=
B_0
\frac{d^2}{z^2+d^2},
\label{eq:single_B_profile}
\end{equation}
where $B_0=B(0)$ is the magnetic-field strength at the lower boundary and $d$ determines the characteristic spatial scale of the magnetic-field inhomogeneity. Magnetic-field extrapolations show that elementary flux tubes generally expand with height as the field strength decreases away from the photospheric footpoints. In particular, the comparison performed by \citet{Demoulin_1994} between observed coronal loops and magnetic flux tubes reconstructed from field extrapolations showed that the latter exhibit a systematic expansion with height, although the predicted expansion is generally stronger than that inferred from the observed loop widths. The profile in Equation~\eqref{eq:single_B_profile} is not intended to reproduce the detailed magnetic structure of an individual coronal loop, but rather to provide a smooth and analytically tractable model of this generic expansion. To specify the stochastic heating at the lower boundary, we adopt the same distribution of heating-event temperatures used in our previous work \citep{barbieri2026,barbieri2026b}. The probability density for an event to heat the plasma to a temperature $T$ is taken to be
\begin{equation}
    \gamma(T)
    =
    \frac{1}{\Delta T}
    \exp\left(
    -\frac{T-T_0}{\Delta T}
    \right),
    \qquad T>T_0,
    \label{eq:heating_temperature_distribution}
\end{equation}
where $\Delta T$ sets mean temperature increase associated with this distribution. The exponential form of $\gamma(T)$ represents a heating scenario in which relatively weak events are considerably more frequent than events capable of heating the plasma to coronal temperatures. This choice is motivated by observations of small-scale brightenings, which indicate that most heating events remain confined to transition-region temperatures, whereas only a small fraction reach temperatures of order $10^6\,{\rm K}$ \citep{Huang2023,Dolliou2023,Dolliou_2024,Dolliou2025}. As discussed by \cite{Barbieri2025c}, the resulting density and temperature profiles depend only weakly on the specific functional form adopted for $\gamma(T)$, provided that its mean temperature is kept fixed and remains of the order of coronal temperatures, i.e., \(\left\langle T-T_0\right\rangle = \Delta T \sim 10^6\,{\rm K}.\) 

\begin{figure*}[t]
    \centering
    \includegraphics[width=\textwidth]{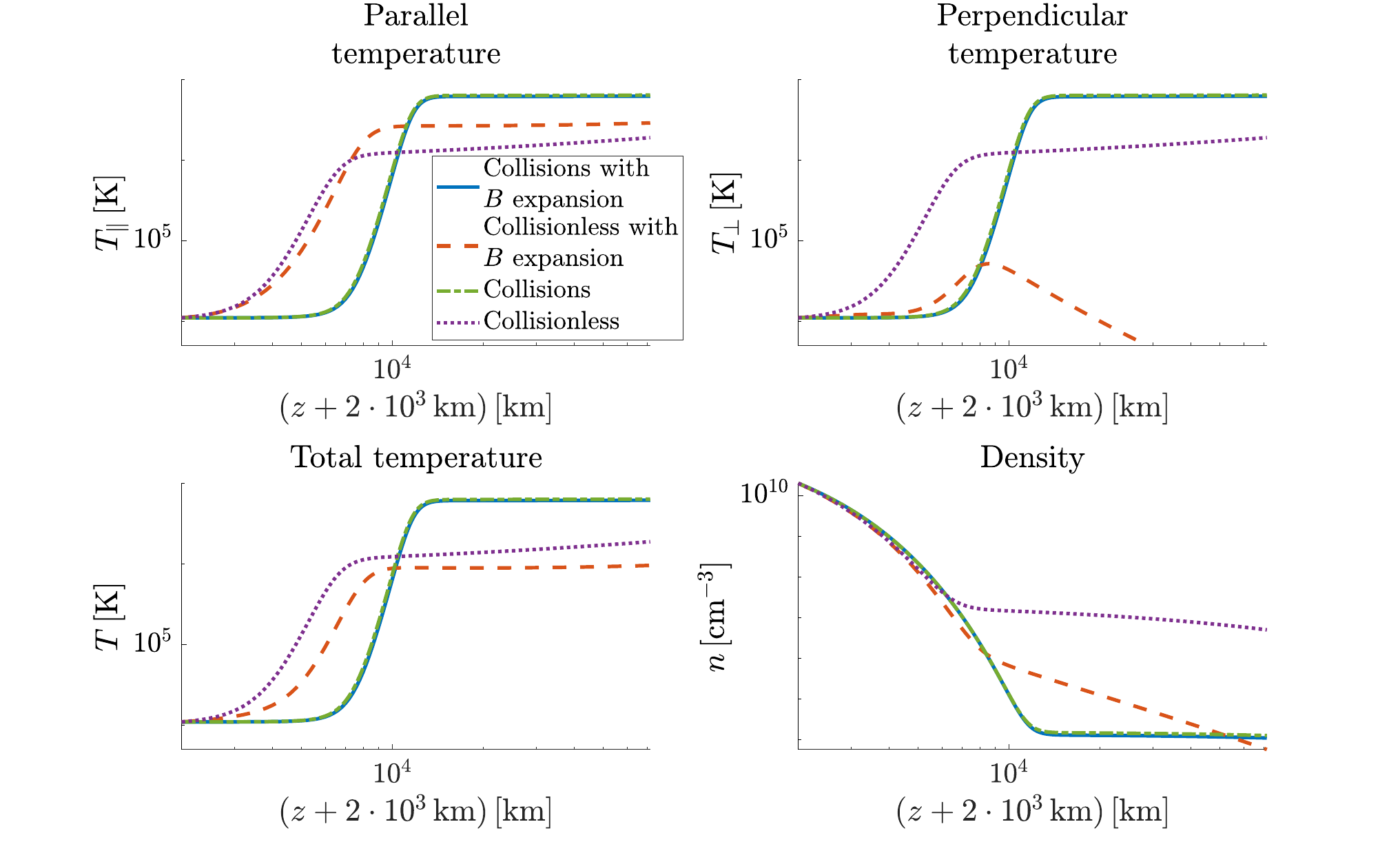}
    \caption{
    Comparison of the temperature and density profiles obtained for the
    different kinetic models considered in this work and in our previous
    studies \citep{barbieri2024temperature,Barbieri2024b,barbieri2026,barbieri2026b}. The panels show the
    parallel temperature $T_{\parallel}$, the perpendicular temperature
    $T_{\perp}$, the total temperature $T$, and the number density $n$
    as functions of height. The four curves
    show the results of our present model including both Coulomb collisions
    and magnetic-field inhomogeneity (solid blue), the collisionless model with
    magnetic-field inhomogeneity (dashed red), the collisional model without
    field inhomogeneity (dash-dotted green), and the purely collisionless
    model without field inhomogeneity (dotted purple), as indicated in
    the legend. The parameters used in
    the calculation are $n_0=2\times10^{10}\,{\rm cm^{-3}}$, $T_0=10^4\,{\rm K}$,
    $d=2\times10^3\,{\rm km}$, and $\Delta T=10^6\,{\rm K}$, with the
    heating-event temperature distribution given by Equation~\eqref{eq:heating_temperature_distribution}.
    }
    \label{fig:figure1}
\end{figure*}

\subsection{Results}

Figure~\ref{fig:figure1} compares the density and temperature profiles predicted by the present model, which accounts for gravitational filtering, Coulomb collisions, and magnetic-field
inhomogeneity, with those obtained in three limiting cases: the collisionless model with constant flux-tube cross section \citep{barbieri2024temperature,Barbieri2024b},
the collisional model with constant flux-tube cross section \citep{barbieri2026}, and the collisionless model with magnetic-field inhomogeneity \citep{barbieri2026b}. The density is computed with Equation~\eqref{eq:collisional_density}, while the parallel and perpendicular temperatures are obtained with
Equation~\eqref{eq:collisional_parallel_temperature}. The total temperature is then computed with Equation~\eqref{eq:collisional_scalar_temperature}. We obtain the collisionless limit
 by setting $T_M(s)=T_0$ and
$\omega(T,s)=R(T,s)=1$ for all $s$, whereas we obtain the constant-cross-section limit by setting $c(s)=0$ for all $s$. The purely collisionless constant-cross-section case is therefore the result of applying both limits simultaneously. 

In the purely collisionless constant-cross-section case (dotted purple curves in Figure~\ref{fig:figure1}), the temperature inversion results solely from gravitational filtering of the multi-temperature distribution
generated by stochastic heating at the lower boundary. The different temperature components have different gravitational scale heights, so that the relative contribution of the hotter populations progressively increases with height. Since the magnetic field is constant, the distribution remains isotropic, with $T_\parallel=T_\perp$, while the total temperature increases toward
coronal values. When Coulomb collisions are included while the flux-tube cross section is kept constant (dash-dotted green curves in Figure~\ref{fig:figure1}), collisional relaxation progressively reduces the contribution of the original suprathermal populations to the local maxwellian core. The collisional efficiency depends strongly on particle energy: lower-energy suprathermal particles are more efficiently thermalized by collisions, whereas the more energetic particles have longer collisional lengths and travel over larger distances before being incorporated into the Maxwellian core. This selective attenuation explains both the lower density and the higher coronal temperature with respect to the purely collisionless constant-cross-section case. A substantial fraction of the particles that would reach coronal heights in the collisionless model are thermalized at lower altitudes, thereby reducing the coronal density. At the same time, the suprathermal population that reaches larger heights is increasingly dominated by hotter components, which are less affected by collisions. The relative weight of high-energy
particles in the coronal velocity distribution is therefore enhanced, leading to a higher coronal temperature despite the reduction in the number of particles reaching these heights.

A different behaviour emerges when magnetic-field inhomogeneity is introduced in the collisionless limit (dashed red curves in Figure~\ref{fig:figure1}). Gravitational filtering still preferentially selects the hotter components because of their larger gravitational scale heights. As the magnetic-field strength decreases along the flux tube,
conservation of the magnetic moment leads to a reduction in the perpendicular kinetic energy, while conservation of total particle energy determines the corresponding variation of the
parallel kinetic energy in the presence of the gravitational and Pannekoek potentials. Together with the restriction of the velocity-space domain connected to the lower boundary, these effects produce a pronounced temperature anisotropy. In particular, the decrease in magnetic-field strength, combined with the pitch-angle-dependent connectivity condition, leads to a much stronger depletion of the perpendicular pressure than of the parallel pressure, resulting in $T_\parallel > T_\perp$. The restriction of the collisionlessly accessible velocity space also produces an additional density depletion relative to the collisionless constant-cross-section case. The present calculation shows that these strong signatures of magnetic-field inhomogeneity are almost completely suppressed once Coulomb collisions are included (solid blue curves in Figure~\ref{fig:figure1}). The profiles obtained with both collisions and magnetic-field inhomogeneity are nearly indistinguishable from those of the collisional constant-cross-section
model. This behaviour results from the efficient collisional redistribution of particles in velocity space. Collisional pitch-angle scattering progressively populates the region that is disconnected from the lower boundary in the collisionless mapping, thereby suppressing the anisotropy associated with the inhomogeneous magnetic field. In terms of the reduced closure adopted here, as $R(T,s)$ evolves toward $R(T,s)\simeq0$, the completed geometrical factors satisfy $\mathcal{G}_i(T,s)\rightarrow1$, and the distinction between the parallel and perpendicular geometrical moments is progressively erased. Consequently, for realistic parameters considered here, collisional pitch-angle redistribution is sufficiently efficient that $T_\parallel$ and $T_\perp$ remain nearly equal throughout the
transition region and corona despite the expansion of the magnetic flux tube. Finally, we emphasize that the reference value of $d$
adopted here corresponds to a particularly strong magnetic-field inhomogeneity, representing the most extreme
configuration considered in the previous collisionless study \citep{barbieri2026b}. We have also varied $d$ by one order of magnitude above and below the reference value used in Figure~\ref{fig:figure1}. Throughout this range, the density and temperature profiles obtained with both Coulomb collisions and magnetic-field inhomogeneity remain nearly indistinguishable from those obtained with collisions alone. This result demonstrate that, under the plasma conditions considered here, the collisional suppression of the anisotropy associated with magnetic-field
inhomogeneity is robust even for substantial variations in the magnetic-field geometry.

To summarize, the comparison of the four cases therefore shows that gravitational filtering remains responsible for the temperature inversion. Coulomb collisions strongly modify the transport of the heated populations and efficiently suppress the additional density depletion and temperature anisotropy produced by the magnetic-field inhomogeneity.

\begin{figure*}
    \centering
    \includegraphics[width=\textwidth]{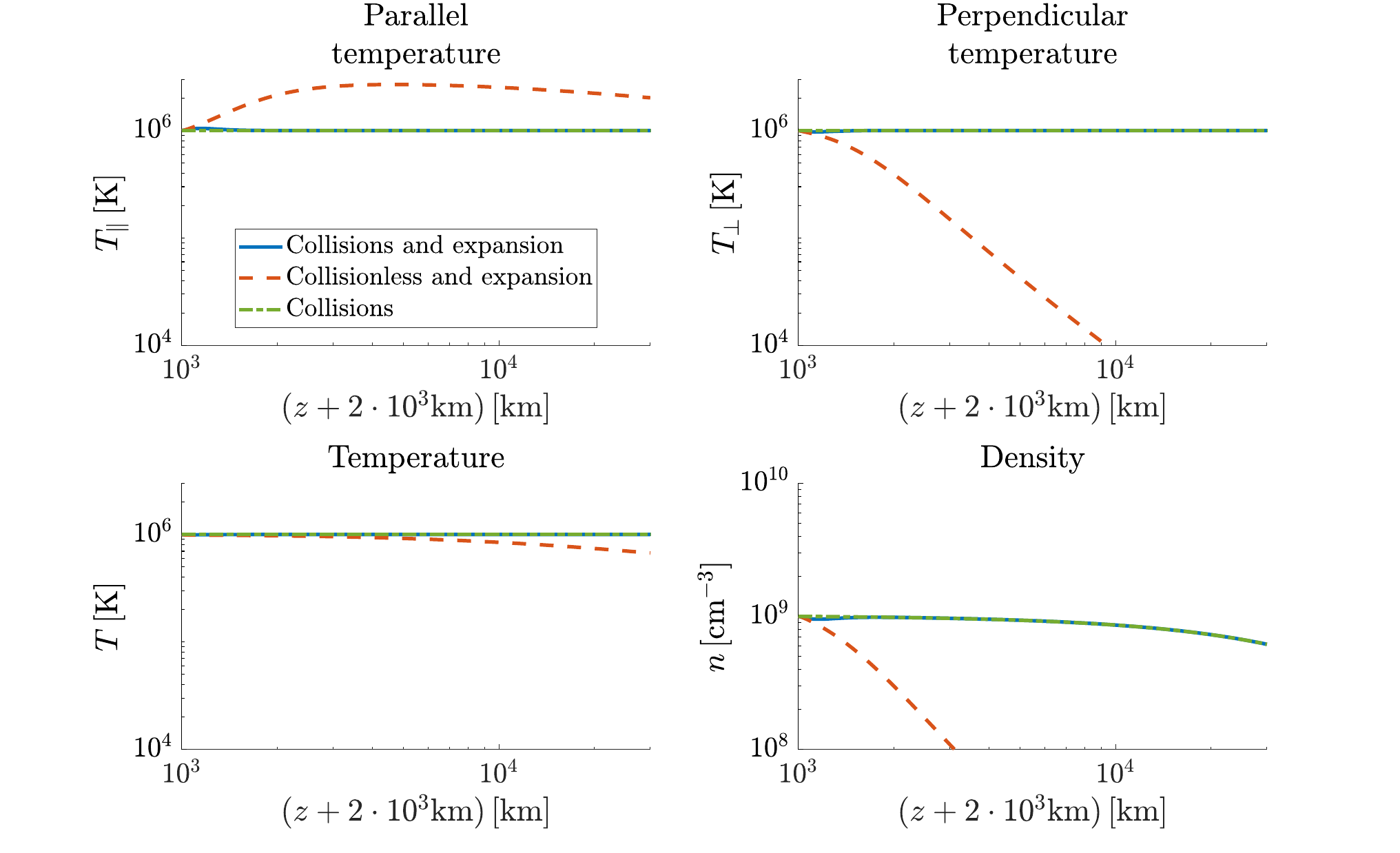}
    \caption{
    The same quantities as in Figure~\ref{fig:figure1} are shown, but for the long-time-scale regime, in which the heated population is described by a single Maxwellian.
    The figure compares the collisional solution with magnetic field inhomogeneity (solid blue curves), the collisionless solution with magnetic field inhomogeneity (dashed red curves), and the collisional constant-cross-section solution (dashed dotted green curves). The parameters adopted in the calculation are the same of Figure \ref{fig:figure1} apart from $n_0=10^{9}\,{\rm cm^{-3}}$, $\hat{T}=10^6$~K. 
    }
    \label{fig:figure2}
\end{figure*}

\section{Long-time-scale regime}\label{sec:long_time_scale_regime}

We now consider the long-time-scale regime, in which the characteristic duration $\tau$ and waiting time $t_w$ of the heating events are comparable to or longer than the proton crossing time $t_{c,p}$
\begin{equation}
    \tau,t_w \gtrsim t_{c,p}.
\end{equation}
As analysed by \citet{barbieri2026}, in this regime, an individual heating event persists sufficiently long for the plasma to mix within a heated flux tube. The lower-boundary distribution of a heated loop can therefore be described by a single Maxwellian at a temperature $\widehat{T}$, and the corresponding loop evolves toward an isothermal configuration.

\subsection{Single-temperature limit of the multi-temperature model}

Within the formalism developed in the present work, the single-temperature regime can be obtained directly as limit of the multi-temperature distribution introduced in Equation~\eqref{eq:halo_distribution_collisional} by setting the heating-temperature distribution to
\begin{equation}
    \gamma(T)=\delta(T-\widehat{T}),
    \label{eq:single_temperature_gamma}
\end{equation}
so that the integral over the different heated populations collapses onto a single Maxwellian population at temperature $\widehat{T}$. The density and the parallel and perpendicular pressure moments are then evaluated from the same expressions used in the multi-temperature model, namely Equations~\eqref{eq:collisional_density} and \eqref{eq:collisional_parallel_pressure}, with all temperature-dependent quantities evaluated at $T=\widehat{T}$. In the long-time-scale regime, as anticipated above, each individual heated loop is approximately isothermal, in contrast to the fast-time-scale regime. The temperature inversion discussed by \citet{barbieri2026} therefore does not develop within individual loops, as it does in the fast-time-scale regime, but emerges only after spatial averaging over an ensemble of unresolved flux tubes characterized by different temperatures. The resulting temperature and density profiles are compatible with the observed ones \citep{barbieri2026}.

\subsection{Strongly collisional single-temperature regime}

An important property of the long-time-scale regime is its high collisionality. As shown by \cite{barbieri2026}, for the realistic parameters considered here, the plasma within an individual heated loop is strongly collisional. We adopt a base density $n_0\simeq10^9\,{\rm cm}^{-3}$, representative of the density inferred for small-scale coronal brightenings and campfire-like events \citep{Berghmans:2021wl,Zhukov2021}, together with a mean temperature $\widehat{T}\simeq10^6\,{\rm K}$. For these parameters, the collisional mean free path is much smaller than the macroscopic length scale of the loop. At the lower boundary, this corresponds to a Knudsen number of order
\begin{equation}
    {\rm Kn}_0
    =
    \frac{\lambda(\widehat{T},0)}{L}
    \sim 10^{-3},
\end{equation}
where $L=0.1R_{\odot}$ denotes the typical vertical size of the system and $R_{\odot}$ is the solar radius. The plasma is therefore well within the strongly collisional regime. As an order-of-magnitude estimate of the transition to collisionless behaviour, we characterise the place where ${\rm Kn}_0\sim1$, i.e., the location where the collisional length becomes comparable to the macroscopic length $L$. From Equation~\eqref{eq:halo_mean_free_path}, at fixed temperature and to leading order in the density dependence, ${\rm Kn}_0$ scales approximately as $n_0^{-1}$, apart from the dependence contained in the Coulomb logarithm. Starting from
${\rm Kn}_0\sim10^{-3}$ at $n_0=10^9\,{\rm cm}^{-3}$, the condition ${\rm Kn}_0\sim1$ is therefore reached, to order of magnitude, when
\begin{equation}
    n_{0,{\rm Kn}\sim1}
    \sim
    10^6\,{\rm cm}^{-3}.
    \label{eq:critical_density_kn}
\end{equation}
Thus, for $\widehat{T}\simeq10^6\,{\rm K}$ and $L=0.1R_{\odot}$, the density adopted here is approximately three orders of magnitude larger than the density at the location where ${\rm Kn}_0\sim1$. Under these conditions, Coulomb collisions efficiently thermalize the particle population over distances much shorter than the characteristic length of the loop. As found in the constant-cross-section model of \cite{barbieri2026}, an individual heated population remains close to a Maxwellian characterized by the single temperature $\widehat{T}$, and the corresponding temperature profile remains approximately isothermal along the loop.

\subsection{Effect of magnetic-field inhomogeneity}

Figure~\ref{fig:figure2} illustrates the effects of magnetic-field inhomogeneity and Coulomb collisions in the single-temperature configuration, obtained by adopting Equation~\eqref{eq:single_temperature_gamma} with $T=\widehat{T}$. As in Figure~\ref{fig:figure1}, we compare the profiles obtained with and without collisions and magnetic-field inhomogeneity. In the collisionless limit, magnetic-field inhomogeneity produces effects similar to those discussed in the fast-time-scale regime (dashed red curves). The restriction of the velocity-space domain connected to the lower boundary leads to a pronounced
temperature anisotropy, with $T_\parallel>T_\perp$, and an additional density depletion. Consequently, the total temperature departs from the isothermal profile, despite the single-temperature Maxwellian prescribed at the boundary.

For the realistic plasma parameters considered here, Coulomb collisions largely suppress these effects (solid blue curves). The resulting temperature and density profiles nearly coincide with those of the collisional constant-cross-section model, with $T_\parallel\simeq T_\perp\simeq\widehat{T}$. The extent of this suppression, however, depends on
the plasma collisionality. This result follows directly from the strong collisionality quantified above. For ${\rm Kn}_0\sim10^{-3}$, collisional redistribution operates over a spatial scale much shorter than the characteristic macroscopic
scale of the loop. Collisional pitch-angle scattering therefore efficiently counteracts the anisotropy associated with the collisionless accessibility condition and progressively populates the region of velocity space that would otherwise remain disconnected from the lower boundary. We conclude that the inclusion of an inhomogeneous magnetic field  does not alter the main physical picture obtained by \cite{barbieri2026} even for the long-time-scale regime. As in the fast-time-scale regime, we have verified that the results are insensitive to variations of the parameter $d$ by one order of magnitude above and below the reference value adopted in Figure~\ref{fig:figure2}.

\subsection{Mechanical equilibrium of the reduced single-temperature solution}

A final point of discussion concerns the mechanical equilibrium of the reduced single-temperature solution. When magnetic-field inhomogeneity and the collisional filling of the disconnected velocity-space domain are simultaneously included, imposing a fixed temperature $\widehat{T}$ does not, in general, provide an exact solution of the parallel pressure-balance equation, Equation~\eqref{eq:anisotropic_pressure_balance}. The reason is that the spatial evolution of $R(\widehat{T},s)$ modifies the velocity moments along the loop and therefore introduces an additional contribution to their spatial derivatives. As a result, the prescribed isothermal solution can leave a finite residual in the parallel momentum balance during the intermediate stage of the collisional evolution. As shown analytically in Appendix~\ref{app:single_temperature_balance}, this residual vanishes in the collisionless limit and in the limit in which the disconnected velocity-space domain is completely filled, but is in general non-zero between these two limiting regimes. For the strongly collisional parameters adopted in Figure~\ref{fig:figure2}, $R(\widehat{T},s)$ decreases rapidly toward zero with height. The distribution then approaches the isotropic single-temperature limit, for which the pressure-balance residual vanishes.

\section{Summary, Limitations and Perspectives}
\label{sec:summary}

In this work, we investigate the combined effects of stochastic heating, gravitational filtering, magnetic-field inhomogeneity, and Coulomb collisions in a coronal magnetic flux tube. Here, we extend our previous collisionless study \citep{barbieri2026b} by including Coulomb collisions and investigate how collisional redistribution modifies the kinetic effects produced by magnetic-field inhomogeneity and gravitational filtering. Collisions are incorporated through a reduced prescription describing both the progressive thermalization of suprathermal populations and the pitch-angle
redistribution of particles into the velocity-space region that is disconnected from the lower boundary in the collisionless mapping. Coulomb collisions strongly suppress the kinetic signatures of magnetic-field inhomogeneity for the realistic plasma parameters considered here. In the collisionless limit, the conservation of particle energy and magnetic moment restricts the accessible region of velocity space. The resulting pitch-angle-dependent depletion produces a substantial density reduction and a pronounced temperature anisotropy, with $T_\parallel>T_\perp$. Once collisions are included, collisional pitch-angle scattering progressively populates the disconnected velocity-space region and reduces the difference between the parallel and perpendicular pressures. The resulting macroscopic profiles therefore approach those obtained in the collisional constant-cross-section model of \cite{barbieri2026}.

In the short-time-scale regime, this collisional isotropization does not remove the temperature inversion generated by stochastic heating. Populations produced at different temperatures coexist within the same flux tube, and gravitational filtering continues to select preferentially the hotter components at increasing height. Magnetic-field inhomogeneity effects on the density and temperature anisotropy are largely suppressed when Coulomb collisions are included at the level considered here. The resulting profiles are therefore predominantly controlled by gravitational filtering and collisional thermalization.

The suppression of the impact of magnetic-field inhomogeneity is still present in the long-time-scale regime. Individual heated loops are highly collisional and approach a single-temperature configuration. In the collisionless model, the pitch-angle-dependent restriction of the velocity-space domain connected to the lower boundary would produce a strongly anisotropic distribution, even when the prescribed boundary Maxwellian is isotropic. Coulomb collisions instead redistribute particles in velocity space, yielding an approximately isotropic distribution with $T_\parallel\simeq T_\perp\simeq\widehat{T}$. The approximately isothermal profiles obtained for individual heated loops in our previous long-time-scale model is therefore recovered even when magnetic-field inhomogeneity is taken into account. In this regime, temperature inversion emerges only after spatial averaging over an ensemble of unresolved loops characterized by different temperatures, yielding temperature and density profiles with values comparable to those observed in the solar corona \cite{Barbieri2025c}.

The reduced collisional prescription adopted here has some limitations. The evolution of the attenuation factor $\omega(T,s)$ was rigorously derived in our previous constant-cross-section model \citep{barbieri2026}. By contrast, the evolution of $R(T,s)$, which describes the collisional filling of the velocity-space domain disconnected from the lower boundary, is introduced here phenomenologically. In particular, the assumption that both processes are characterized by the same collisional length $\lambda(T,s)$ is a simplifying hypothesis and does not follow directly from the Coulomb collision operator. Indeed, energy-space relaxation and pitch-angle diffusion involve distinct contributions to the Fokker--Planck collision operator and need not be characterized by identical spatial scales. A self-consistent determination of the filling of the disconnected domain would require solving the collisional kinetic equation for the velocity distribution function, including pitch-angle diffusion across the collisionless connectivity boundary. Such a treatment lies beyond the scope of the present model. The present prescription is conceptually related to closures adopted in exospheric kinetic models, in which velocity-space domains not directly connected to the lower boundary may be assumed to be fully populated, depleted, or partially populated \citep{Pierrard1996,PIERRARD2001,Pierrard2008,PtersdeBonhome2025}. This analogy is limited to the treatment of velocity-space completion
and does not extend to the choice of the equation that governs the
evolution of the collisional filling function $R$.

As in our previous work \citep{barbieri2026}, our treatment of collisional
relaxation relies on the approximation that each initially
Maxwellian population retains its shape in velocity space,
while its moments are progressively reduced through the
temperature-dependent survival factor $\omega(T,s)$.
This approximation neglects the explicit dependence of
the Coulomb collision frequency on the individual particle
velocity, which scales approximately as $v^{-3}$ for
suprathermal particles. Consequently, even within a single
Maxwellian population initially characterized by temperature
$T$, particles in the thermal core undergo collisional
relaxation more rapidly than those in the high-energy tail.
A velocity-dependent survival factor would therefore
progressively distort the initial Maxwellian distribution,
preferentially preserving its suprathermal tail.
By assigning the same attenuation factor to all particles
within a given temperature component, our model tends to
overestimate the collisional relaxation of the most energetic
particles. We therefore expect our model to underestimate the coronal density. Nevertheless, we
expect the qualitative behaviour of the density and
temperature profiles to remain unchanged. A fully kinetic description
of this velocity-dependent relaxation would require solving
the Coulomb Fokker--Planck collision operator, for example
using the Rosenbluth potentials,
which would introduce substantially greater computational
complexity and is beyond the scope of the present model.

Results of present study also indicate a clear direction for the extension of the present framework. The model considered here assumes a symmetric loop with no field-aligned bulk flow and velocity distribution
functions that are symmetric with respect to the parallel velocity. Relaxing this assumption would allow for asymmetric stochastic heating at the two footpoints to generate asymmetric velocity distribution functions and a finite field-aligned particle flux and heat-flux. Such configurations can produce siphon flows \citep[e.g.][]{Reale2010} along coronal loops and require the ambipolar electric field to be determined self-consistently together with the electron and proton distribution functions.

\section*{Acknowledgments}
\textbf{The authors wish to thank the anonymous reviewer for the valuable comments and suggestions that helped improve the clarity and quality of this work.} L.B. wants to thank Sorbonne Universit\'e for financial support within the framework of the Initiative Physique des Infinis. L.B. wants to thank Filippo Pantellini for useful discussion. D.V. is funded by the Deutsche Forschungsgemeinschaft (DFG, German Research Foundation) -- Project number 568092272.

\bibliographystyle{aasjournalv7}
   \bibliography{manuscript} 

\appendix

\section{Evolution of the Maxwellian core}
\label{app:core_evolution}

In this Appendix, we provide the physical interpretation of the terms entering the evolution equations for the amplitude $C(s)$, that is Eq. \eqref{eq:C_evolution}, and
temperature $T_M(s)$ of the Maxwellian core, that is Eq. \eqref{eq:TM_evolution_compact}.

\subsection{Evolution of the core amplitude}
\label{app:C_terms}

The evolution of the core amplitude $C(s)$ follows from the local particle-conservation condition introduced in Equation~\eqref{eq:density_collisional_variation}. The different terms entering its evolution describe the redistribution of particles associated with the collisional attenuation of the suprathermal populations and with the progressive filling of the velocity-space domains that are disconnected from the lower boundary under the assumption of  collisionless trajectories. The first contribution inside the temperature integral,
\begin{equation}
    \frac{\mathcal{G}_*(T,s)}{\lambda(T,s)},
    \nonumber
\end{equation}
describes the reduction of the density carried by each suprathermal Maxwellian population through collisional thermalization. As the survival factor $\omega(T,s)$ decreases, particles are removed from
the corresponding suprathermal population and transferred to the Maxwellian core. The second contribution,
\begin{equation}
    -
    \frac{
        R(T,s)\left[1-G_*(T,s)\right]
    }{
        \lambda(T,s)
    },
    \nonumber
\end{equation}
arises from the simultaneous collisional filling of the disconnected velocity-space domain. As $R(T,s)$ decreases, this domain becomes progressively populated through collisional pitch-angle scattering. The associated increase in the density carried by the suprathermal population partially compensates the density reduction produced by its collisional thermalization and therefore reduces the corresponding increase required in the core amplitude $C(s)$ by local particle conservation. Finally, the contribution
\begin{equation}
    \frac{
        C(s) R_M(s)
        \left[1-G_{*,M}(s)\right]
    }{
        \mathcal{G}_{*,M}(s)\lambda_M(s)
    },
    \nonumber
\end{equation}
accounts for the analogous collisional filling of the disconnected velocity-space domain associated with the Maxwellian core. As $R_M(s)$ decreases, an increasing fraction of the local velocity space becomes populated by the core. At fixed $C(s)$, this filling would increase the core density.

\subsection{Evolution of the core temperature}
\label{app:TM_terms}

We now describe the terms entering the evolution equation for the core temperature $T_M(s)$ given by Equation~\eqref{eq:TM_evolution_compact}. The term
\begin{equation}
    -A_t\frac{{\rm d}\mathcal{H}_\parallel}{{\rm d}s}
    \nonumber
\end{equation}
contains the complete spatial variation of the suprathermal parallel pressure. Since $\mathcal{H}_\parallel$ depends on the Boltzmann factor, the collisional survival factor $\omega(T,s)$, and the completed geometrical factor $\mathcal{G}_\parallel(T,s)$, it includes the effects of the potential, collisional thermalization, magnetic-field inhomogeneity, and the filling of the disconnected velocity-space domain. The contribution
\begin{equation}
-
\frac{1}{k_{\rm B}}
\frac{{\rm d}\phi}{{\rm d}s}
\left(
A_t\mathcal{H}_*
+
C E_M\mathcal{G}_{*,M}
\right)
\nonumber
\end{equation}
is associated with the force, whereas
\begin{equation}
    \Delta\mathcal{P}
    \frac{{\rm d}\ln B}{{\rm d}s}
    \nonumber
\end{equation}
is the geometrical contribution associated with the anisotropic pressure tensor and the variation of the magnetic-field strength. The latter vanishes for an isotropic distribution or for a constant
magnetic-field strength. The term
\begin{equation}
-
\frac{d C}{d s}\,
T_M E_M\mathcal{G}_{\parallel,M}
\nonumber
\end{equation}
accounts for the variation of the core parallel pressure associated with the evolution of its amplitude. The remaining contribution,
\begin{equation}
-
C T_M E_M
\left[
\left.
\frac{d\mathcal{G}_{\parallel,M}}
{d s}
\right|_{T_M}
-
\frac{\mathcal{G}_{\parallel,M}}
{k_{\rm B}T_M}
\frac{{\rm d}\phi}{{\rm d}s}
\right],
\nonumber
\end{equation}
contains the explicit spatial variation of the Maxwellian-core parallel pressure at fixed $T_M$. The derivative $(d\mathcal{G}_{\parallel,M}/d s)_{T_M}$ accounts for the spatial variation of the velocity-space geometry and of the disconnected-domain filling at fixed core temperature, whereas the second term originates from the explicit spatial derivative of the Boltzmann factor $E_M$. Finally, the denominator
\begin{equation}
C E_M
\left[
\mathcal{G}_{\parallel,M}
\left(
1+
\frac{\phi}{k_{\rm B}T_M}
\right)
+
T_M
\frac{d\mathcal{G}_{\parallel,M}}
{d T_M}
\right]
\end{equation}
represents the response of the Maxwellian-core parallel pressure to a variation of $T_M$, with the remaining variables held fixed. The derivative $d\mathcal{G}_{\parallel,M}/d T_M$ accounts for the change in the velocity-space contribution of the Maxwellian core as its temperature varies.

\section{Numerical solution of the collisional system}
\label{app:numerical_scheme}

In this Appendix, we describe the numerical procedure used to solve the collisional system introduced in the main text. The method follows
the numerical strategy of \citet{barbieri2026}, extended here to account for magnetic-field inhomogeneity and the collisional filling of the trapped velocity-space domain. The continuous temperature dependence of the suprathermal population is first replaced by a finite temperature grid. The resulting problem can then be written as a finite system of coupled ordinary differential equations along the curvilinear coordinate $s$.

\subsection{Temperature discretization}
\label{app:temperature_discretization}

The temperature integrals are truncated at a sufficiently large temperature $T_{\max}$ such that the contribution of the tail of $\gamma(T)$ above $T_{\max}$ is negligible. We introduce a uniform
temperature grid
\begin{equation}
    T_j
    =
    T_0+(j-1)\Delta T,
    \qquad
    j=1,\ldots,N_T,
\end{equation}
with
\begin{equation}
    \Delta T
    =
    \frac{T_{\max}-T_0}{N_T-1}.
\end{equation}
For a generic function $F(T,s)$, the temperature integral is evaluated using the trapezoidal rule,
\begin{equation}
    \int_{T_0}^{T_{\max}}
    F(T,s)\,{\rm d}T
    \simeq
    \sum_{j=1}^{N_T}
    q_j F(T_j,s),
    \label{eq:app_trapezoidal_rule}
\end{equation}
where the quadrature weights are
\begin{equation}
    q_1=q_{N_T}=\frac{\Delta T}{2},
    \qquad
    q_j=\Delta T,
    \quad
    j=2,\ldots,N_T-1.
\end{equation}

At each temperature node, we define
\begin{equation}
    \omega_j(s)
    \equiv
    \omega(T_j,s).
\end{equation}
The discretized collisional evolution equation is therefore
\begin{equation}
    \frac{{\rm d}\omega_j}{{\rm d}s}
    =
    -\frac{\omega_j(s)}{\lambda_j(s)}, \quad \lambda_j(s)
    \equiv
    \lambda(T_j,s),
    \quad
    j=1,\ldots,N_T,
    \label{eq:app_omega_j}
\end{equation}

Since $R$ obeys the same equation with the same boundary condition, its discrete values are obtained directly as
\begin{equation}
    R_j(s)
    \equiv
    R(T_j,s)
    =
    \omega_j(s).
    \label{eq:app_Rj}
\end{equation}
Thus, no additional differential equations for $R$ need to be solved.

\subsection{Discrete evaluation of the velocity moments}
\label{app:discrete_moments}

At every position $s$, the collisionally completed geometrical factors are evaluated at each temperature node as
\begin{equation}
    \mathcal{G}_{i,j}(s)
    =
    1-
    R_j(s)
    \left[
    1-G_i(T_j,s)
    \right],
    \qquad
    i\in\{*,\parallel,\perp\}.
    \label{eq:app_Gij}
\end{equation}

The suprathermal contribution to the density integral is
\begin{equation}
S_*(s) \equiv
\int_{T_0}^{T_{\max}} {\rm d}T\,
\gamma(T)\,\omega(T,s)\,
e^{-\phi(s)/(k_{\rm B}T)}\,\mathcal{G}_*(T,s)
\simeq
\sum_{j=1}^{N_T} q_j\,\gamma(T_j)\,\omega_j(s)\,
e^{-\phi(s)/(k_{\rm B}T_j)}\,\mathcal{G}_{*,j}(s).
\label{eq:app_S0}
\end{equation}
$R_M(s)$ is not evolved independently. Instead, it is obtained from
Equation~\eqref{eq:RM_evolution}. Since the instantaneous value of $T_M(s)$ does not in general coincide with one of the temperature nodes $T_j$, the value of $R_M(s)$ is evaluated numerically by interpolating the discrete collisional solution $\{R_j(s)\}=\{\omega_j(s)\}$ at the current value of
$T_M(s)$. We then evaluate the corresponding core geometrical factors  as
\begin{equation}
    \mathcal{G}_{i,M}(s)
    =
    1-
    R_M(s)
    \left[
    1-G_i(T_M(s),s)
    \right],
    \qquad
    i\in\{*,\parallel,\perp\}.
    \label{eq:app_GiM}
\end{equation}

We reconstruct the macroscopic quantities  algebraically from the current numerical solution as
\begin{equation}
    n(s)
    =
    n_0
    \left[
    A_t S_*(s)
    +
    C(s)E_M(s)\mathcal{G}_{*,M}(s)
    \right],
    \label{eq:app_density}
\end{equation}
\begin{equation}
    p_{\parallel}(s)
    =
    n_0k_{\rm B}
    \left[
    A_t S_\parallel(s)
    +
    C(s)T_M(s)E_M(s)
    \mathcal{G}_{\parallel,M}(s)
    \right],
    \label{eq:app_pparallel}
\end{equation}
and
\begin{equation}
    p_{\perp}(s)
    =
    n_0k_{\rm B}
    \left[
    A_t S_\perp(s)
    +
    C(s)T_M(s)E_M(s)
    \mathcal{G}_{\perp,M}(s)
    \right].
    \label{eq:app_pperp}
\end{equation}

The corresponding local parallel, perpendicular, and scalar temperatures are computed from Equations~\eqref{eq:collisional_parallel_temperature}
and \eqref{eq:collisional_scalar_temperature}, respectively.

\subsection{Evaluation of the collisional mean free paths}
\label{app:mean_free_paths}

Once the macroscopic quantities are reconstructed, we evaluate the Coulomb logarithm independently at each temperature node as
\begin{equation}
    \ln\Lambda_j(s)
    =
    \ln\left[
    \frac{3k_{\rm B}\, T_j}{2e^2}
    \sqrt{
    \frac{k_{\rm B}\, T(s)}
    {4\pi e^2 n(s)}
    }
    \right].
    \label{eq:app_coulomb_log_j}
\end{equation}

The second-order velocity moments associated with the $j$-th Maxwellian population are
\begin{equation}
    \left\langle
    v_{\parallel}^2
    \right\rangle_j
    =
    \frac{k_{\rm B}\, T_j}{m}
    \frac{
    \mathcal{G}_{\parallel,j}(s)
    }{
    \mathcal{G}_{*,j}(s)
    },
    \label{eq:app_vparallel_j}
\end{equation}
and
\begin{equation}
    \left\langle
    v_{\perp}^2
    \right\rangle_j
    =
    \frac{k_{\rm B}\, T_j}{m}
    \frac{
    \mathcal{G}_{\perp,j}(s)
    }{
    \mathcal{G}_{*,j}(s)
    }.
    \label{eq:app_vperp_j}
\end{equation}
Here, $\langle v_{\perp}^2\rangle_j$ refers to one of the two perpendicular velocity components. The corresponding mean free path is then
\begin{equation}
    \lambda_j(s)
    =
    \frac{
    m^2
    \left[
    \left\langle v_{\parallel}^2\right\rangle_j
    +
    2\left\langle v_{\perp}^2\right\rangle_j
    \right]^2
    }{
    4\pi e^4 n(s)\ln\Lambda_j(s)
    }.
    \label{eq:app_lambda_j}
\end{equation}

We use the same prescription to evaluate the mean free path associated with the Maxwellian core.

\subsection{Discrete equations for $C(s)$ and $T_M(s)$}
\label{app:C_TM_discrete}

We evaluate the temperature integral entering the evolution equation for the Maxwellian-core amplitude using the same quadrature approach. Defining
\begin{equation}
I_C(s) =
\int_{T_0}^{T_{\max}} {\rm d}T\,
\gamma(T)\,e^{-\phi(s)/(k_{\rm B}T)}\,\omega(T,s)
\left[
\frac{\mathcal{G}_*(T,s)}{\lambda(T,s)}
-\frac{R(T,s)[1-G_*(T,s)]}{\lambda(T,s)}
\right].
\label{eq:app_IC_cont}
\end{equation}
we obtain
\begin{equation}
I_C(s) \simeq
\sum_{j=1}^{N_T} q_j\,\gamma(T_j)\,
e^{-\phi(s)/(k_{\rm B}T_j)}\,\omega_j(s)
\left[
\frac{\mathcal{G}_{*,j}(s)}{\lambda_j(s)}
-\frac{R_j(s)[1-G_*(T_j,s)]}{\lambda_j(s)}
\right].
\label{eq:app_IC}
\end{equation}

Since $R_j=\omega_j$, we are able to evaluate this expression entirely from the single set of evolved collisional variables $\{\omega_j\}$. The evolution equation for the Maxwellian-core amplitude is then
\begin{equation}
\frac{{\rm d}C}{{\rm d}s}
= \frac{A_t\,I_C(s)}{E_M(s)\,\mathcal{G}_{*,M}(s)}
- \frac{C(s)\,R_M(s)[1-G_{*,M}(s)]}
{\mathcal{G}_{*,M}(s)\,\lambda_M(s)}.
\label{eq:app_C_discrete}
\end{equation}

For compactness, we write this equation as
\begin{equation}
    \frac{{\rm d}C}{{\rm d}s}
    =
    F_C
    \left(
    s,\{\omega_j\},C,T_M
    \right),
    \label{eq:app_FC}
\end{equation}
where the required values of $R_j$ and $R_M$ follow algebraically from Equations~\eqref{eq:app_Rj} and interpolation, respectively. Similarly, the quantities entering the suprathermal contribution to the anisotropic pressure balance are 
\begin{equation}
    \mathcal{H}_*(s)
    \simeq
    \sum_{j=1}^{N_T}
    q_j\,
    \gamma(T_j)\,\omega_j(s)\,
    e^{-\phi(s)/(k_{\rm B}\, T_j)}\,
    \mathcal{G}_{0,j}(s)
\end{equation}
and
\begin{equation}
    \mathcal{H}_i(s)
    \simeq
    \sum_{j=1}^{N_T}
    q_j\, T_j\,
    \gamma(T_j)\,\omega_j(s)\,
    e^{-\phi(s)/(k_{\rm B}\, T_j)}\,
    \mathcal{G}_{i,j}(s),  \quad
    i\in\{\parallel,\perp\}.
\end{equation}

The evolution equation for the Maxwellian-core temperature is then evaluated from Equation~\eqref{eq:TM_evolution_compact}, using the discrete
moments and the current values of the local plasma quantities. We write
\begin{equation}
    \frac{{\rm d}T_M}{{\rm d}s}
    =
    F_T
    \left(
    s,\{\omega_j\},C,T_M
    \right).
    \label{eq:app_FT}
\end{equation}
All trapped-domain $R$ required in $F_T$ are reconstructed from the same numerical solution for $\omega$ and are therefore not
additional dynamical variables.

\subsection{Reduction to a system of ordinary differential equations}
\label{app:ODE_system}

After temperature discretization, we reduce the original integro-differential problem to a finite-dimensional system of ordinary differential equations. Since we obtain $R_j=\omega_j$ and $R_M$ 
algebraically from the temperature interpolation, the state vector contains only the $N_T$ collisional variables $\omega_j$, together
with the Maxwellian-core amplitude and temperature,
\begin{equation}
    \mathbf{y}(s)
    =
    \left(
    \omega_1,\ldots,\omega_{N_T},
    C,T_M
    \right)^{\rm T}.
    \label{eq:app_state_vector}
\end{equation}

The dimension of the differential system is therefore $N_{\rm ODE} = N_T+2$. The complete numerical problem is written as
\begin{equation}
    \frac{{\rm d}\mathbf{y}}{{\rm d}s}
    =
    \mathbf{F}
    \left(
    s,\mathbf{y}
    \right).
    \label{eq:app_ODE_system}
\end{equation}

The first $N_T$ components of $\mathbf{F}$ are
\begin{equation}
    F_j
    =
    -\frac{\omega_j}{\lambda_j},
    \qquad
    j=1,\ldots,N_T,
\end{equation}
while the final two components are
\begin{equation}
    F_{N_T+1}
    =
    F_C(s,\mathbf{y}),
\end{equation}
and
\begin{equation}
    F_{N_T+2}
    =
    F_T(s,\mathbf{y}).
\end{equation}

The boundary conditions are
\begin{equation}
    \omega_j(0)=1,
    \qquad
    j=1,\ldots,N_T,
\end{equation}
together with Eq. \eqref{eq:core_boundary_conditions}.

\subsection{Stiffness and implicit integration}
\label{app:implicit_solver}

The finite-dimensional system in Equation~\eqref{eq:app_ODE_system} is intrinsically stiff. Because there is a wide separation of collisional scales between low-energy
and suprathermal particles and is not specific to the present reduced description. A direct kinetic treatment based on the full Fokker--Planck collision operator would need to resolve the same
multi-scale collisional dynamics in velocity space, leading to an even more demanding numerical problem. An important advantage of
the present approach is that the collisional dynamics is reduced to a one-dimensional system of ordinary differential equations along $s$, so that the resulting stiffness can be efficiently handled using an adaptive implicit ODE solver. The origin of the stiffness in the present formulation can be understood directly from the collisional relaxation equations,
\begin{equation}
    \frac{{\rm d}\omega_j}{{\rm d}s}
    =
    -\frac{\omega_j}{\lambda_j}.
    \label{eq:app_generic_relaxation}
\end{equation}

The Coulomb mean free path depends strongly on particle energy. Consequently, the coldest Maxwellian components easily have $\lambda_j$ much smaller than the characteristic macroscopic length
scales over which the density, temperature, and magnetic-field strength vary. At the same time, the hottest populations easily possess much larger mean free paths and therefore evolve over macroscopic distances. The complete integration would consequently be forced to follow the microscopic mean free path of the coldest populations even in regions where all macroscopic quantities vary over much larger spatial scales. An explicit scheme is therefore numerically inefficient for the present multi-scale problem. Following \citet{barbieri2026}, we instead integrate the system using an adaptive implicit solver suitable for stiff ordinary differential
equations. In the present implementation, we use MATLAB \texttt{ode15s}. The implicit treatment allows us to follow the rapidly relaxing
collisional modes without imposing a globally
microscopic spatial step, while the adaptive step-size control automatically increases the integration step when the solution evolves
on larger spatial scales. We check the numerical solution through standard convergence and consistency tests. In particular, we increase $T_{\max}$ and $N_T$ until no appreciable changes are observed in the density and
temperature profiles. The numerical scheme also recovers the collisionless solution when all collisional mean free paths become much larger than the characteristic size of the system.

\section{Pressure balance in the single-temperature limit}
\label{app:single_temperature_balance}

In this Appendix, we examine whether the reduced kinetic model satisfies the stationary pressure-balance equation in the single-temperature limit introduced in Sec. \ref{sec:long_time_scale_regime}. We consider a population at a fixed temperature $\widehat{T}$ so that the density and pressure components are
\begin{equation}
    n(s)
    =
    n_0
    \exp\left[-\frac{\phi(s)}{k_B\widehat{T}}\right]
    \mathcal{G}_*(\widehat{T},s)
    \label{eq:app_single_n}
\end{equation}
and
\begin{equation}
    p_i(s)
    =
    n_0 k_B\widehat{T}
    \exp\left[-\frac{\phi(s)}{k_B\widehat{T}}\right]
    \mathcal{G}_i(\widehat{T},s),
    \qquad i\in\{\parallel,\perp\}.
    \label{eq:app_single_p}
\end{equation}

To determine whether the single-temperature solution satisfies the pressure-balance relationship in Equation~\eqref{eq:anisotropic_pressure_balance}, we define the pressure-balance residual
\begin{equation}
    R_P
    \equiv
    \frac{dp_\parallel}{ds}
    -
    (p_\parallel-p_\perp)
    \frac{d\ln B}{ds}
    +
    n\frac{d\phi}{ds}.
    \label{eq:app_pressure_residual}
\end{equation}
Equilibrium conditions correspond to
$R_p=0$. Using Equations~\eqref{eq:app_single_n},
\eqref{eq:app_single_p}, and \eqref{eq:G_completed_general}, the residual becomes
\begin{equation}
    R_P
    =
    -
    n_0 k_B\widehat{T}
    \exp\left[-\frac{\phi}{k_B\widehat{T}}\right]
    \frac{dR}{ds}
    (1-G_\parallel).
    \label{eq:app_residual_final}
\end{equation}

Equation~\eqref{eq:app_residual_final} shows explicitly that the departure from total pressure balance is associated with the spatial evolution of $R$. In particular, the single-temperature model satisfies the pressure-balance equation exactly whenever $R$ is spatially constant. In the collisionless limit, $R=1$ everywhere and therefore $dR/ds=0$. Equation~\eqref{eq:app_residual_final} immediately gives $R_p=0$, showing that the collisionless single-temperature solution is an exact
stationary solution of the parallel momentum equation, including the effects of magnetic-field inhomogeneity. The equilibrium equation is also recovered in the absence of magnetic-field inhomogeneity. For a constant-cross-section flux tube, $B(s)=B_0$. Finally, the same result is recovered in the strongly collisional limit, where $R=0$. In this case, Equation~\eqref{eq:app_residual_final} immediately yields $R_P=0$. A finite residual arises only when the magnetic-field inhomogeneity and a spatially evolving $R$ are simultaneously present. This residual is therefore a specific consequence of the reduced model adopted to describe the collisional population of the velocity-space domain that is disconnected from the lower boundary in the collisionless mapping.

\end{document}